\documentclass[12pt,a4paper,spanish]{article}
\usepackage[ansinew]{inputenc}
\usepackage{babel}
\usepackage{latexsym}
\usepackage{amsmath}
\usepackage{amsfonts}
\usepackage{amssymb}
\usepackage{textcomp}
\usepackage{graphicx}
\usepackage{enumerate}

\begin{document}

\begin{titlepage}
\begin{center}
{\large Dynamical variables in Gauge-Translational Gravity}

\vglue 1cm

J. JULVE* and A. TIEMBLO$^+$

{\it *$^+$Instituto de Física Fundamental, CSIC, \break C/Serrano
113 bis, Madrid 28006, Spain

*julve@iff.csic.es

 $^+$tiemblo@iff.csic.es}

\end{center}

\vglue 1cm

\noindent {Assuming} that the natural gauge group of gravity is
given by the group of isometries of a given space, for a maximally
symmetric space we derive a model in which gravity is essentially a
gauge theory of translations. Starting from first principles we
verify that a nonlinear realization of the symmetry provides the
general structure of this gauge theory, leading to a simple choice
of dynamical variables of the gravity field corresponding, at first
order, to a diagonal matrix, whereas the non-diagonal elements
contribute only to higher orders.

\vglue 0.5cm


\vglue 1cm

\begin{center}
\end{center}

\vglue 3cm

\noindent{*Corresponding author.}

\end{titlepage}

\section{Introduction}

The establishment of a gauge symmetry lies on the empirical evidence
of a locally invariant property related to a group of
transformations. In this sense the existence of a continuous
ten-parameter group (Poincar\'e) giving rise to the conservation of
the fundamental dynamical variables, strongly suggests the existence
of a relevant link between dynamics and the basic properties of the
space-time i.e geometry.

From the geometrical point of view Poincar\'e can be defined as the
group of isometries of Minkowsky space, so that, being essentially
gravity a dynamical theory of the space-time, it seems natural to
consider the group of isometries of a given space as the gauge group
of such a theory.

Minkowsky is the simplest case (zero curvature) of a maximally 3+1
symmetric space and thus it excludes the presence of a cosmological
constant. On the other hand, our knowledge of the geometrical
properties of the space-time is only phenomenological and therefore
it is approximate. Strictly speaking what we observe is that the
space is approximately homogeneous and isotropic, and that it is
endowed with the kinematical Lorentz group of Relativity so that we
can assume that the symmetry group of space-time seems to be very
close to Poincar\'e. Consequently we assume that the general
candidate for a gauge theory of gravity is the group of isometries
of a maximally symmetric space (the limit of zero curvature being
the Poincar\'e group).

Lastly, the evidence that elementary matter is fermionic strongly
supports the hypothesis that gravity couples to it through the
vierbein. The assumption that the vierbein is the connection of the
local translations makes it to transform as a tensor under
diffeomorphisms and under the (even local) Lorentz group. These
properties are obtained by defining the vierbeins as a NLR of
Poincar\'e group (cosets with respect to lorentz).

The need to couple the fermionic matter to gravity stems from the
attempts to enlarge the geometrical framework of General Relativity
with the introduction of a suitable internal group
\cite{Utiyama}\cite{Sciama}\cite{Kibble}\cite{Hayashi}\cite{Ivanenko}\cite{Lord1}\cite{Lord2}\cite{Sardanashvily}\cite{Ashtekar}.

The search for an unified description seems to suggest, as
reasonable starting point, the adoption of a common and general
gauge scheme for all interactions including the gravity itself.
These ideas gave raise to the programm of finding approaches in
which gravity is mediated by gauge connections as it happens for the
remaining fundamental forces \cite{Hehl1}\cite{Hehl2}\cite{Hehl3}.
The appearance of tetrads, an object with holonomic and nonholonomic
indices, is, to this purpose, an unavoidable requirement. We claim
that tetrads are the fingerprint of the presence of translations in
the Gauge Group, a natural feature in a theory like gravity which
can be essentially considered as a dynamical theory of the
space-time itself. As we have mentioned in a previous paper this
reminds the Feynman's words "Gravity is that field which corresponds
to a gauge invariance with respect the displacement
transformations". We stress that, as it shall be shown in what
follows, the natural way to realize a symmetry containing
translations is precisely a non linear realization where the cosets
have the form $e^{{\rm i}p_i\varphi^i}$, where the set of fields
$\varphi^i$, which becomes isomorphic to the coordinates, acts as
the parameters which characterize the coset. In this way the fields
$\varphi^i$ introduce a dynamical interpretation of an ingredient
like the coordinates which is present in any field theory. On the
other hand,it can be seen that, from the group theoretical point of
view, they behave as the Goldstone bosons with respect the gauged
translations.

To make the paper as self contained as possible we include in
Section 2 a brief review sketching the general lines of the non
linear local realizations of the space-time groups. In Section 3 we
deduce, starting from first principles and definitions, the
integrability conditions which determine the structure of the gauge
theory, serving, at the same time, as a link between the gauge and
the geometrical description. Section 4 is devoted to establish the
structure of the gauge theory which provides us the underlying
background of the canonical geometrical description, allowing, for
instance, alternative and simpler choices of the dynamical
variables, an essential question in gravity theories. We conclude
with some final remarks on the possible extensions and open
problems.

\section{The structure of the tetrads}

We briefly review here some fundamental tools and results from
previous works.

A maximally four dimensional symmetric space admits a maximal number
of Killing vectors supporting a semisimple Lie algebra described by
the ten generators:

\begin{equation}
P_i={\rm i}\,\{\partial_i+\frac{k}{4}(2x_ix^j-\delta_i^j
r^2)\partial_j\}
\end{equation}
\begin{equation}
L_{ij}={\rm i}\,(\delta_i^k x_j - \delta_j^k x_i)\partial_k
\end{equation}

\noindent{where} $r^2=\eta_{i j}x^i x^j$ and being $k$ the sectional
curvature. The commutation relations can be written in the form:

\begin{equation}
[P_i,P_j]={\rm i}\,k L_{i j}
\end{equation}
\begin{equation}
[L_{i j},P_k]={\rm i}\,\eta_{k[i} P_{j]}
\end{equation}
\begin{equation}
[L_{ij},L_{kl}]=-{\rm i}\,\{\eta_{i[k}L_{l]j}-\eta_{j[k}L_{l]i}\}
\end{equation}

\noindent{which} reduces to Poincar\'e when $k\rightarrow0$. The
occurrence of "translational-like" transformations rises the problem
of realizing a local symmetry of this kind in which the Lorentz
subgroup $H$ still be linearly represented, as dictated by the
particle phenomenology. The natural choice is given by a local non
linear realization with cosets defined as $ e^{{\rm i}\,\varphi^i
P_i}$ \cite{Coleman}\cite{Callan}\cite{Salam}\cite{Isham}, which is
the most general one preserving the linear action of the subgroup
$H$.

The non linear gauge realizations of space-time symmetry groups
containing translations have been the object of several papers
\cite{Borisov}\cite{Stelle}\cite{Julve1}\cite{Julve2}\cite{Lopezpinto}\cite{Lopezpinto2}\cite{Tresguerres}
in the past. Nevertheless, in order to make this work more readable
we include here a brief review of the methods and main results.

Let $G$ be a Lie group having a subgroup $H$, we assume that the
elements $C(\varphi)$ (cosets) of the quotient space $G/H$ can be
characterized by a set of parameters say $\varphi$. Let us denote by
$\psi$ an arbitrary linear representation of the subgroup $H$.

The non linear realization can be derived from the action of a
general element  $"g"$ of the whole group on the coset
representatives defined in the form:

\begin{equation}\label{1}
g\,C(\varphi)=C(\varphi')h(\varphi,g)
\end{equation}
where
$h(\varphi,g)\in H$. It acts linearly on the representation space
$\psi$ according to:

\begin{equation}\label{2}
\Psi'=\varrho[h(\varphi,g)]\,\Psi,
\end{equation}
being $\varrho[h]$ a representation of the subgroup $H$.

The next step to construct a non linear local theory is to define
suitable gauge connections. They can be obtained by substituting the
ordinary Cartan 1-form $\omega=C^{-1}dC$ by a generalized expression
of the form:

\begin{equation}\label{3}
\Gamma=C^{-1}DC
\end{equation}
where $D=d+\Omega$ is the covariant differential built with the
1-form connection $\Omega$ defined on the algebra of the whole group
and having the canonical transformation law:

\begin{equation}
\Omega'=g\Omega g^{-1}+  gdg^{-1}
\end{equation}

The generalized local Cartan 1-form is:
\begin{equation}
\Gamma=C^{-1}{\bf D} C=e^{-{\rm i}\,\varphi^iP_i}(d+{\rm
i}\,T^iP_i+\frac{\rm i}{2}A^{ij}L_{ij})e^{{\rm i}\,\varphi^iP_i}
\end{equation}
where $T$ is the linear translational connection and $A^{ij}$ the
corresponding one for the Lorentz group.

Using Hausdorff-Campbell formulas to deal with exponentials,  after
a little algebra (the details can be found in
\cite{Martin1}\cite{Martin2}) we obtain:
\begin{equation}
\Gamma={\rm i}\,{\hat e}^iP_i+\frac{i}{2}{\hat A}^{ij}L_{ij}
\end{equation}
where ${\hat e}^i$ and ${\hat A}^{ij}$ are the 1-form non linear
local connections given by the following expressions:
\begin{equation}
{\hat
e}^i=ND\varphi^i+\frac{1-N}{\mu^2}(\varphi^jD\varphi_j)\varphi^i+M\,T^i+\frac{1-M}{\mu^2}(T^j\varphi_j)\varphi^i
\end{equation}
and
\begin{equation}
{\hat
A}^{ij}=A^{ij}+\frac{1-M}{\mu^2}\varphi^{[i}D\varphi^{j]}+k\,N\varphi^{[i}T^{j]}
\end{equation}
where $D\varphi^i\equiv(d\varphi^i+A^i_j\varphi^j)$ is the Lorentz
covariant differential,
$\mu^2\equiv\eta_{ij}\,\varphi^i\varphi^j\equiv\varphi_i\varphi^i$,
and $M$ and $N$ are given by the following series:
\begin{equation}
M=1-\frac{k\mu^2}{2!}+\frac{(k\mu^2)^2}{4!}+...\sim {\rm
cos}\sqrt{k\mu^2}
\end{equation}
and
\begin{equation}
N=1-\frac{k\mu^2}{3!}+\frac{(k\mu^2)^2}{5!}+...\sim\frac{1}{\sqrt{k\mu^2}}\;{\rm
sin}\sqrt{k\mu^2}
\end{equation}

In (10), the translational connection 1-form $T^i$ has dimensions of
length. In order to have a dimensionless connection $\gamma^i$
homogeneous with the ordinary Lorentz connection $A^{ij}$, we
introduce a constant characteristic length, say $\lambda$\,, and
define $T^i=\lambda\gamma^i$. We notice that the occurrence of a
fundamental length is a common feature in gravity theories
\cite{Borzeskowski}\cite{Garay}\cite{Berg}\cite{Feinberg}\cite{Kato}\cite{Konishi},
as for instance the Planck scale, at the basis of string theory, or
the spacing parameter in lattice theories. In our scheme, we claim
that a characteristic length finds its natural place in the
translational connection above. Then the smallness of $\lambda$
emphasizes the interpretation of gravity as a perturbation of a
background (usually supposed flat) metric.

In order to express all the objects in terms of only the non linear
connections we introduce in (12) the value of $A_{ij}$ worked out
from (13) in terms of ${\hat A}_{ij}\,$,  obtaining
\begin{equation}
{\hat e}^i=\frac{N}{M}{\hat
D}\varphi^i+\frac{1}{\mu^2}(1-\frac{N}{M})(\varphi^jd\varphi_j)\varphi^i+\lambda(
\frac{1}{M}\,{\bar\gamma}^i+\frac{\varphi^i\varphi_j}{\mu^2}\gamma^j)
\end{equation}
where $\hat D$ stands for the Lorentz covariant differential in
terms of ${\hat A}_{ij}$\,, and
${\bar\gamma}^i=(\delta_{ij}-\frac{\varphi^i\varphi_j}{\mu^2})$.

Our focus now is on the structure of the non-linear vierbein (16).
In its limit for $\lambda\rightarrow0$ one has
\begin{equation}
{\hat e}(0)^i=\frac{N}{M}{\hat
D}\varphi^i+\frac{1}{\mu^2}(1-\frac{N}{M})(\varphi^jd\varphi_j)\varphi^i\;.
\end{equation}
whereas in the limit $k\rightarrow0$ we have $M=N=1$, so that it
becomes
\begin{equation}
e^i=D\varphi^i+\lambda\gamma^i\;\equiv e(0)^i+\lambda\gamma^i\;,
\end{equation}
which is the expression for the Poincar\'e case, where $e^i$ and its
non-linear version ${\hat e}^i$ coincide.

Now we remind that the essential feature of a tetrad is given by its
double character, transforming as general vector in the Greek
indices and as a Lorentz vector in the Latin ones, so that it
provides a link between both spaces.

Now we observe that the covariant derivative of a Lorentz vector
like ${\hat e}(0)_\mu^i$ in (17) and (18) is the minimal structure
able to take the role of a tetrad, so we shall call it "minimal
tetrad". We stress that the difference between ${\hat e}(0)_\mu^i$
and ${\hat e}_\mu^i$ regards the behavior under local translations,
due to the presence of the connection $\gamma_\mu^i\,$.

\section{Integrability conditions}

The passage to the geometrical description can be made with the help
of the general tetrad (16) defining, as usual, a metric tensor of
the form:
\begin{equation}
g_{\mu\nu}={\hat e}_{i\mu}{\hat e}^i_\nu\,
 =g(0)_{\mu \nu}+\lambda\gamma_{(\mu
 \nu)}+\lambda^2\gamma_{\mu \rho}\gamma_{\nu\sigma}\,g(0)^{\rho
 \sigma}\;,
\end{equation}
where $g(0)_{\mu \nu}={\hat e}(0)_{i\mu}{\hat e}(0)_\nu^i$ is the
corresponding "minimal metric tensor", and we have used ${\hat
e}(0)_{i\mu}$ and its formal inverse ${\hat e}(0)^\nu_j$ to
transform indices.

Two comments are now in order. The first one concerns equation(19)
that imitates a weak field expansion over a background metric
$g(0)_{\mu \nu}$. It must be emphasized however that it is not a
perturbation approach but an exact result derived from the
underlying gauge structure, which is apparent only at the vierbein
level. Secondly, the decomposition (16) implies a non-trivial
structure for the formal inverse ${\hat e}^\mu_i$ present in the
definition of the contravariant metric tensor. We explicitly assume
that the theory is analytical in the characteristic length
$\lambda$, so that the formal inverses are given by an expansion in
powers of $\lambda$. Strictly speaking a similar question arises
with the definition of $g(0)^{\rho \sigma}$ present in (19), and we
are to show that the structure and properties of this minimal metric
tensor can be derived from general integrability conditions.

In a previous work \cite{Martin1} we have seen that the field
equations of gravity in the vacuum can be interpreted as a gauge
theory of translations defined in the metric of a maximally
symmetric background space. Now we are going to show that this
result holds without making recourse to the field equations even in
the presence of matter, or, in other words, as a consequence of the
underlying gauge structure of the theory which is previous to any
dynamics.

The analyticity in $\lambda$ lets to work out the existence
conditions and ensuing properties of the solutions in the limit
$\lambda\rightarrow0$ (minimal tetrads).  To this end we first
redefine the Lorentz connection ${\hat A}_\mu^{ij}$ as follows:
\begin{equation}
{\hat A}_\mu^{ij}={\hat e}^{\alpha i}{\cal D}_\mu {\hat
e}_\alpha^j+B_\mu^{ij},
\end{equation}
where ${\cal D}_\mu$ is the ordinary Christoffel covariant
derivative acting on the coordinate index $\alpha$ of the tetrad
${\hat e}_\alpha^j$ . The first term of this redefinition, usual in
gauge theories of gravity, describes the value of the Lorentz
connection in the absence of matter, whereas the second one
$B_\mu^{ij}$ takes into account the coupling with the spin densities
present in the matter terms, so that $B_\mu^{ij}=0$ in the vacuum.

Now we shall see that the background metric can be derived from
integrability conditions which are previous to the equations of
motion. This requires some rather involved algebra that we briefly
outline in the following.

For $\lambda=0$, contracting (20) with $\varphi^j$ we get
\begin{equation}
{\hat A}_{\mu j}^i\varphi^j=[e(0)^{\alpha i}{\cal
D}(0)_\mu[e(0)_{\alpha j}\varphi^j]-\partial_\mu\varphi^i+B_\mu^i],
\end{equation}
where ${\cal D}(0)_\mu$ is the Christoffel covariant derivative
constructed with the metric tensor $g(0)_{\mu \nu}$ , and
$B_\mu^i=B_{\mu j}^i\varphi^j$. Using (21) and the notation
$\chi\equiv k\,\mu^2$ , (17) becomes
\begin{equation}
{\hat e}(0)^i_\mu=\frac{N}{M}[{\hat e}(0)^{\alpha i}
\frac{1}{2k}{\cal D}(0)_\mu {\cal
D}(0)_\alpha\chi+B_\mu^i]+(1-\frac{N}{M})\frac{1}{2k}{\cal
D}(0)_\mu\chi\,\varphi^i
\end{equation}
Contracting it with ${\hat e}(0)_{i\nu}$ one obtains
\begin{equation}
g(0)_{\mu\nu}=\frac{1}{2k}\frac{N}{M}[{\cal D}(0)_\mu {\cal
D}(0)_\nu\chi + (\frac{N}{M}-1)\frac{1}{2k}{\cal D}(0)_\mu\chi {\cal
D}(0)_\nu\chi]+\frac{N}{M}B_{\mu\nu}
\end{equation}
The square bracket in (23)can be brought to the form $H(\chi){\cal
D}(0)_\mu {\cal D}(0)_\nu F(\chi)$. Exploiting the explicit values
(14) and (15), the function $F$ and the integration factor $H$ turn
out to be $M$ and $(cN)^{-1}$ respectively, where $c$ is a constant.
With $c=(2k)^{-1}$ we obtain
\begin{equation}
g(0)_{\mu\nu}=\frac{1}{M}\,{\cal D}(0)_\mu {\cal D}(0)_\nu \,M +
\frac{N}{M}\,B_{\mu\nu}
\end{equation}
Taking the trace of (24) we have:
\begin{equation}
M=\frac{1}{4}\,(\Box(0)M+B)\;,
\end{equation}
where $B$ is the trace of $B_{\mu\nu}$ and $\Box(0)$ is the
covariant d'Alembertian corresponding to $g(0)_{\mu\nu}$ .
Substituting in (24) we finally obtain
\begin{equation}
D(0)_\mu D(0)_\nu M=\frac{1}{4}\,\Box(0)M-{\bar B}_{\mu\nu},
\end{equation}
being ${\bar B}_{\mu\nu}\equiv
B_{\mu\nu}-\frac{1}{4}\,g(0)_{\mu\nu}B$.

Taking now symmetric and antisymmetric parts of (26) we get:
\begin{equation}
D(0)_\mu D(0)_\nu M
=\frac{1}{4}\,g(0)_{\mu\nu}\Box(0)M-\frac{1}{2}\,{\bar
B}_{(\mu\nu)},
\end{equation}
and
\begin{equation}
{\bar B}_{[\mu\nu]}=B_{[\mu\nu]}=0 \;,
\end{equation}
which in Lorentz indices gives
\begin{equation}
B_{[\mu\nu]}=e(0)_{[\mu}^k e(0)_{\nu]}^i B_{kij}\varphi^j=0 \;\;\;\;
  \Rightarrow \;\;\;\; B_{[ki]j}\varphi^j=0
\end{equation}

We recall that the fields $\varphi^j$ are by definition independent
functions as long as, being the Goldstone bosons of the gauged
translations, there are not dynamical relations among them.
Otherwise stated, their motion equations are satisfied identically
since they yield the null covariant divergence of the Einstein
tensor. Therefore (29) is verified only when $B_{[ki]j}=0$ . Taking
into account the antisymmetry of $B_{kij}$ in the last two indices
we can write
\begin{equation}
B_{kij}=B_{ikj}=-B_{ijk},
\end{equation}
so that symmetrizing in $i j$ one gets finally
\begin{equation}
B_{(ij)k}=0 \;.
\end{equation}
As a consequence, at least at zeroth order in $\lambda$ , equation
(27) reduces to
\begin{equation}
{\cal D}(0)_\mu{\cal D}(0)_\nu\,M= \frac{1}{4}\,g(0)_{\mu\nu}\Box
(0)M\;.
\end{equation}

It must be emphasized here that the question is not finding a
solution $M$ to (32), which we know {\it a priori}, but
acknowledging that its mere existence implies, as a well known
integrability condition, the maximally symmetric character of the
space. Thus $g(0)_{\mu\nu}$ is determined from first principles and
previously to any dynamics. As a particular case, for
$k\rightarrow0$ equation (32) is replaced by
\cite{Martin1}\cite{Martin2}
\begin{equation}
g(0)_{\mu\nu}={\cal D}(0)_\mu {\cal D}(0)_\nu\;\sigma\quad\quad
(\sigma\equiv\frac{1}{2}\mu^2)\;,
\end{equation}
which leads us to a Minkowskian metric. As we shall see, it suffices
to adopt $\varphi^i$ as coordinates to verify that $g(0)_{\mu\nu}$
reduces to the flat metric $\eta_{ij}$ .

Consequently the geometrical description is given in terms of the
finite expansion (19), which depends on the translational connection
(the true gravitational dynamical variable) $\gamma_{\mu\nu}$ and a
background maximally symmetric metric tensor $g(0)_{\mu\nu}\,$.

\section{Gauge structure and dynamical variables}

In this scheme, the dynamical gravitational variables in the
geometrical approach are embodied in the translational connection
$\gamma_{\mu\nu}$ defined on a background metric $g(0)_{\mu\nu}$.
Once $g(0)_{\mu\nu}$ has been determined {\it prior} to any dynamics
by integrability conditions, it is of uppermost interest to pinpoint
the structure of the minimal tetrad $e(0)_\mu^i$ associated to it,
which shall provide us with very useful tools for the identification
of the dynamical variables, a fundamental problem in gravity
theories.

The passage from the gauge description to the geometrical one is
canonically accomplished by using (20) for $B_\mu^{ij}=0$ in the
Field Strength Tensor, which becomes:
\begin{equation}
F_{\mu\nu}^{ij}=e_\alpha^i e_\beta^j R^{\alpha\beta}_{\mu\nu}
\end{equation}

Starting from this relationship we first consider the Poincar\'e
case where the cancelation of the Riemann tensor at zero order in
$\lambda$ stems from the integrability conditions, so that, being
$R(0)^{\alpha\beta}_{\mu\nu}=0$, we conclude that also $F(0)_{\mu
\nu}^{ij}=0$ and then $A_\mu^{ij}$ must be a pure gauge. The
structure of such a connection is given by the inhomogeneous part of
the formal variation of a gauge connection, thus we write:
\begin{equation}
A(0)_\mu^{ij}=U^{ik}\partial_\mu U_k^j,
\end{equation}
where $U^{ik}$ is an arbitrary pseudo-orthogonal matrix describing a
general Lorentz transformation. Putting this in the zeroth order of
(18) we obtain:
\begin{equation}
e(0)_\mu^i=\partial_\mu\varphi^i+U^{ik}\partial_\mu
U_{kj}\varphi^j\;,
\end{equation}
so we can write:
\begin{equation}
e(0)_\mu^i=\partial_\mu \varphi^i+
U^{ik}\partial_\mu[U_{kj}\varphi^j] -\partial_\mu
\varphi^i=U^{ik}\partial_\mu [U_{kj}\varphi^j] =
U^{ik}\partial_\mu{\hat\varphi}_k
\end{equation}
where ${\hat\varphi}_k=U_{kj}\varphi^j$. Then the background metric
may be written as follows:
\begin{equation}
g(0)_{\mu\nu}=U^{ik}\partial_\mu{\hat\varphi}_k U^{il}\partial_\nu
{\hat\varphi}_l=\partial_\mu{\hat\varphi}_k
\partial_\nu{\hat\varphi}^k
\end{equation}

It is immediate to check that (38) satisfies the condition (33). In
a non linear realization of the Poincar\'e group the fields
$\varphi^i$ transform as the cartesian coordinates, thus in a
Minkowskian space they can be properly used as coordinates.

To reproduce the usual geometrical approach we note that the
translational connection $\gamma_{\mu\nu}=e_\mu^i\gamma_{ij} e_\nu^
j$ exhibits an underlying invariance under Lorentz transformations
$U$ in the Latin indexes, so that we can fix the gauge to render
$\gamma_{ij}$ (and consequently $\gamma_{\mu\nu}$) symmetrical, thus
recovering the usual ten degrees of freedom of canonical gravity.

Now we recover the expression (19) of the general metric tensor,
taking the symmetric and antisymmetric parts of $\gamma_{\mu \nu}$:
\begin{equation}
\gamma_{\mu \nu}=\frac{1}{2}\,s_{\mu \nu}+\frac{1}{2}\,a_{\mu \nu },
\end{equation}
being $s_{\mu  \nu}=\gamma_{(\mu \nu)}$ and $a_{\mu
\nu}=\gamma_{[\mu \nu]}$ , so (19) becomes:
\begin{equation}
g_{\mu \nu}=g(0)_{\mu \nu}+\lambda\,s_{\mu
\nu}+\frac{\lambda^2}{4}[s_{\mu\rho}s_{\nu \rho}+s_{(\mu
\rho}a_{\nu) \sigma}+a_{\mu \rho}a_{\nu \sigma}]g(0)^{\rho \sigma}.
\end{equation}
Steering to Lorentz indices, namely $s_{\mu
\nu}=e(0)_\mu^i\,s_{ij}\,e(0)_\nu^j$ and
$a_{\mu\nu}=e(0)_\mu^i\,a_{ij}\,e(0)_\nu^j$ , and adopting the
coordinates $x^\mu$ for the cartesian Goldstone ones $\varphi^i$,
the metric tensor $g(0)_{\mu\nu}$ reduces to $\eta_{i j}$ . Now we
can choose $U$ such that $s_{ij}$ becomes $U^k_is_{kl}U^l_j=d_{ij}$
diagonal, and $a_{ij}\rightarrow U^k_ia_{kl}U^l_j={\hat a}_{ij}$ ,
obtaining:
\begin{equation}
g_{i j}=\eta_{i j}+\lambda\,d_{i j}+\frac{\lambda^2}{4}[d_{i k}d_{j
l}+d_{(i k}\hat{a}_{j)l}+\hat{a}_{i k}\hat{a}_{j l}]\eta^{k l}.
\end{equation}

We then have the usual ten degrees of freedom of canonical gravity,
albeit in quite a different arrangement: the four eigenvalues of the
symmetric part of $\gamma_{\mu\nu}$ and the six elements of an
antisymmetric matrix. We stress that these d.o.f. appear in (41) at
different orders in $\lambda$ so that, for instance, the
calculations at first order get highly simplified.

The case of a maximally symmetric space is slightly more complicated
because $A_\mu^{ij}$ is not a pure gauge. We again start from the
limit $\lambda\rightarrow0$ of (12) that can be written as:
\begin{equation}
e(0)_\mu^i=N
D_\mu\varphi^i+\frac{1-N}{2\chi}\,\partial_\mu\chi\,\varphi^i
\end{equation}
where $\chi\equiv k\mu^2$. Taking into account (37) and using the
relation ${\hat\varphi}^k=U_j^k\varphi^j$ in (42) we obtain:
\begin{equation}
e(0)_\mu^i=U_k^i(N\partial_\mu
{\hat\varphi}^k+\frac{1-N}{2\chi}\,\partial_\mu\chi\,{\hat\varphi}^k)\;.
\end{equation}

A more compact and convenient notation is obtained by redefining the
fields ${\hat\varphi}^k$ according to
\begin{equation}
{\tilde\varphi}^k=F(\chi){\hat\varphi}^k\;,
\end{equation}
which substituted in (43) lead us to
\begin{equation}
e(0)_\mu^
i=U_k^i\frac{N}{F}\,\partial_\mu{\tilde\varphi}^k+(\frac{1-N}{2\chi}-\frac{NF'}{F})\partial_\mu
\chi\,{\tilde\varphi}^k.
\end{equation}

The value of $F$ is chosen so as to cancel the second term on the
right hand side of (45), namely:
\begin{equation}
\frac{1-N}{2x}-\frac{NF'}{F}=0,
\end{equation}
where the prime denotes differentiation with respect to $\chi$.
Using (15) we obtain $ F=\frac{c}{\sqrt{\chi}}\;{\rm
Tan}\frac{\sqrt{\chi}}{2}$ , so that
\begin{equation}
e(0)_\mu^ i=U_k^i\frac{1+M}{c}\;\partial_\mu{\tilde\varphi}^k.
\end{equation}
where $c$ is an integration constant.

Writing now $F(\chi)$ in terms of the redefined fields
${\tilde\varphi}^k$ and choosing $c=2$, the tetrad finally reads
\begin{equation}
e(0)_\mu^i=U_k^i(1+\frac{1}{4}\,{\tilde\chi})^{-1}\partial_\mu\tilde{\varphi}^k
\;.
\end{equation}
Taking again, as in the Poincar\'e case, the Goldstone fields
${\tilde\varphi}^k$ as coordinates, we derive the metric tensor
\begin{equation}
g(0)_{ij}= (1+\frac{1}{4}{\tilde\chi})^{-2}\eta_{ij}\;,
\end{equation}
in which we recognize the so called Riemannian form of the metric
for a space of constant curvature.

\section{ Concluding remarks and first order equations}
The choice of the dynamical variables given in equation (41)
simplifies the structure of the theory. This allows us to find out
the general form and properties of the vacuum equations of gravity
at first order in $\lambda$ . To do this we start from the
Einstein's equations $G_{ij}=\frac{\Lambda}{4}g_{ij}$ in the
presence of a cosmological constant, that can be alternatively
written:
\begin{equation}
R_{ij}+\frac{\Lambda}{4}g_{ij}=0
\end{equation}

There exists,when two different metric tensors $g_{\mu\nu}$ and
$g(0)_{\mu\nu}$ are involved, a useful relation between the
Christophel's connections which highly simplify the calculations,
namely:

\begin{equation}
\Gamma^\rho_{\mu\alpha}=\Gamma(0)^\rho_{\mu\alpha}+\Delta^\rho_{\mu\alpha},
\end{equation}
where $\Gamma^\varrho_{\mu\alpha}$ is the Christophel symbol
constructed with $g_{\mu\nu}$ and $\Gamma(0)^\varrho_{\mu\alpha}$
the corresponding one to $g(0)_{\mu\nu}$, being
\begin{equation}
\Delta^\rho_{\mu\alpha}=\frac{1}{2}g^{\lambda \rho}[D(0)_\mu
g_{\lambda\alpha}+D(0)_\alpha g_{\lambda \mu}-D(0)_\lambda g_{\mu
\alpha}]
\end{equation}
with $D(0)_\mu$ the covariant derivative in terms of
$\Gamma(0)^\rho_{\mu\alpha}$.

According with (41) we are going to use in the following Latin
indexes, being $g(0)_{ij}$ the background metric and
$g_{ij}=g(0)_{ij}+\lambda d_{ij}$ the first order expansion of the
general metric.

The relation between the corresponding Ricci tensors is then given
by the following expression:
\begin{equation}
R_{ij}=R(0)_{ij}+D(0)_j\Delta^k_{ki}-D(0)_k\Delta^k_{ij}+\Delta^l_{ik}\Delta^k_{jl}-\Delta^l_{ij}\Delta^k_{kl},
\end{equation}
where $D(0)_i$ and $R(0)_{ij}$ are respectively the covariant
derivative and the Ricci tensor constructed with the background
metric.

A brief calculation leads to the value of $\Delta^i_{jk}$ which
reads:
\begin{equation}
\Delta^i_{jk}=\frac{\lambda}{2}(D(0)_jd^i_k+D(0)_kd^i_j-D(0)^id_{jk}),
\end{equation}
where the indexes are raised and lowered using the background
metric.

From (56) we see that the zero order terms are satisfied when
$R=-\Lambda$. So that the first order equations becomes:
\begin{equation}
\Box(0)d_{ij}+D(0)_jD(0)_i
d^k_k-D(0)_kD_{(i}d^k_{j)}+\frac{\Lambda}{2}d_{ij}=0,
\end{equation}
and its trace:
\begin{equation}
\Box(0)d^k_k-D(0)_kD(0)_id^{ki}+\frac{\Lambda}{4}d^k_k=0
\end{equation}

It is not the aim of this paper to include a general survey of the
first order solutions, which merit by themselves a more detailed and
specific study. Notwithstanding we are going to briefly comment some
features of the problem relevant in this choice of the dynamical
variables. In fact being $d_{ij}$ a diagonal matrix equation (58)
contains, when $ i\neq j$ additional information with respect to the
usual treatments,namely the second order analytical restrictions:
\begin{equation}
D_k D_{(i}\hat{ d}^k_{j)}=0,
\end{equation}
where $i \neq j$ and:
\begin{equation}
\hat{d}^k_j=d^k_j-\frac{1}{2}\delta^k_j d^l_l.
\end{equation}
Obviously these restrictions are absent in any other choice of the
dynamical variables and gives an important input in the search of
the general scheme of the first order solutions.

Summarizing, our proposal describes the space-time physics by a
twofold assumption, one is the gauge nature of the translations,
which introduces a characteristic length $\lambda$ interpretable as
the "true" gravitational interaction, and the other is the structure
of empty space, attained in the limit $\lambda\rightarrow0$, which
is taken to be a maximally symmetric background space. This
geometrical assumption is equivalent to adopt the existence of a
cosmological constant $\Lambda$ as a phenomenological observation on
the same footing of the approximate planarity (homogeneity and
isotropy) of space-time.

The role played by $\lambda$ in this approach, together with a
principle of economy, strongly suggests to relate this
characteristic length to the gravitational constant. In fact, in
natural units, having a dimensionless vacuum gravity action requires
the introduction of a constant factor with dimensions $L^{-2}$ ,
namely the inverse Newton constant, and it becomes natural its
identification with $\lambda^{-2}$.

Therefore the free gravitational lagrangian $e^\mu _i
e^\nu_jF_{\mu\nu}^{ij}$ should include a depressing factor
$\lambda^2$ with respect to the matter terms. Thus we observe that
when (19) is used, the motion equations stemming from the variation
$\delta A_\mu^{ij}$ become purely algebraical so that $B_{\mu
\nu}^{ij}=0$ in the absence of matter, while $B_\mu^{ij}$ becomes
equal to the matter spin densities coupled linearly to $A_\mu^{ij}$
. Being the matter terms the only ones contributing to the value of
$B_\mu^{ij}$, they are evidently depressed by at least the same
factor $\lambda^2$ existing between the matter Lagrangian and the
vacuum term. This obviously implies that the term $B_{(\mu \nu)}$ do
not contributes to the zeroth order equation (28), the structure of
which agrees with the results of the ordinary dynamical treatment of
gravity as a local field theory. In a gauge field theory only the
fermions give rise to these kind of contributions, and the fermion
spin densities are completely antisymmetric when all indexes are of
the same (either tensor or Lorentz) nature, so that a symmetric part
$B_{(\mu \nu)}$ should be absent in any case.

Therefore we conclude that the identification of the characteristic
length $\lambda$ with the gravitational constant is not only an
economical and natural assumption but, at the same time, fully
consistent with well established theoretical results.

 As it is seen, the appearance of the cosmological constant is, in
this scheme, an initial condition related to the gauge space-time
group considered, so that it constitutes an initial ingredient of
the background space of the theory. From this point of view
revisiting the Quantum Field Theory on maximally symmetric spaces
appears as a very promising topic.

\section{acknowledgements}
We acknowledge Prof. A. Fern\'andez-Rañada and J.
Mart\'{\i}n-Mart\'{\i}n for useful discussions.

\end{document}